\documentclass[nofootinbib, superscriptaddress, reprint, amsmath,amssymb, aps]{revtex4-1}

\usepackage{graphicx}
\usepackage{bm}
\graphicspath{ {./images/} }
\usepackage{braket}

\usepackage[dvipsnames]{xcolor}

\usepackage{footnote}
\usepackage{mathtools}
\usepackage{amsmath}
\usepackage{float}
\usepackage[colorlinks]{hyperref}
\usepackage{dsfont}

\usepackage{enumitem,kantlipsum}

\usepackage{verbatim}

\DeclarePairedDelimiter{\ceil}{\lceil}{\rceil}

\begin{document}

\title{Invariant subspaces of two-qubit quantum gates and their application in the verification of quantum computers}

\author{Yordan S. Yordanov}
\affiliation{ Cavendish Laboratory, Department of Physics, University of Cambridge, Cambridge CB3 0HE, United Kingdom}

\author{Jacob Chevalier-Drori}
\affiliation{ DAMTP, Department of Applied Mathematics and Theoretical Physics, University of Cambridge, Cambridge CB3 0WA, United Kingdom}

\author{Thierry Ferrus}
\affiliation{Hitachi Cambridge Laboratory, Cambridge CB3 0HE, United Kingdom}

\author{Matthew Applegate}
\affiliation{ Cavendish Laboratory, Department of Physics, University of Cambridge, Cambridge CB3 0HE, United Kingdom}

\author{Crispin H. W. Barnes}
\affiliation{ Cavendish Laboratory, Department of Physics, University of Cambridge, Cambridge CB3 0HE, United Kingdom}

\date{\today}

\begin{abstract}

We investigate the groups generated by the sets of $CP$, $CNOT$ and $SWAP^\alpha$ (power-of-SWAP) quantum gate operations acting on $n$ qubits. Isomorphisms to standard groups are found, and using techniques from representation theory, we are able to determine the invariant subspaces of the $n-$qubit Hilbert space under the action of each group. 
For the $CP$ operation, we find isomorphism to the direct product of $n(n-1)/2$ cyclic groups of order $2$, and determine $2^n$ $1$-dimensional invariant subspaces corresponding to the computational state-vectors.
For the $CNOT$ operation, we find isomorphism to the general linear group of an $n$-dimensional space over a field of $2$ elements, $GL(n,2)$, and determine two $1$-dimensional invariant subspaces and one $(2^n-2)$-dimensional invariant subspace. 
For the $SWAP^\alpha$ operation we determine a complex structure of invariant subspaces with varying dimensions and occurrences and present a recursive procedure to construct them. 
As an example of an application for our work, we suggest that these invariant subspaces can be used to construct simple formal verification procedures to assess the operation of quantum computers of arbitrary size.

\end{abstract}

\maketitle

\section{ Introduction}\label{Intro}

% Introduction.
Functional validation is an essential part of all computational development \cite{FuncVal936245, FormVer, VerLang6588543}. Formal verification, where the correspondence between a mathematical model and system output are compared, is a powerful tool employed to prove that a low-level (e.g. a gate array) implementation of a computational system performs its function as intended \cite{FormVer}. At a fundamental level, the complexity of the verification problem in classical computational systems grows exponentially with the number bits - and can therefore be classified as a hard problem \cite{FuncVal936245}. The same is true for quantum computers, though the problem is considerably compounded by the fact that just prior to measurement, qubits can be in a superposition of computational basis states so the output state can only be inferred statistically.

As tangible quantum computers are now on the cusp of practical use \cite{NISQ, NISQ_2, google_supreme, Rigetti, IBM}, there is a growing requirement for formal methodologies regarding their verification to be developed. This requirement is complicated by the fact that there are no quantum computers available that could be used as a reference, it is therefore natural to endeavour to employ classical computers as the solution \cite{sim_quant_comp_1, sim_quant_comp_2, google_supreme}.  However, the use of classical computers quickly becomes infeasible as the size of the quantum computer increases. 
For example, for even a relatively small quantum computer with say 50 qubits, the wave function would require 16 petabytes of data storage. Manipulating such a large amount of data is cumbersome and expensive and few have tried it in this context \cite{google_supreme, multi_qubit_sim_1, multi_qubit_sim_2}. The most powerful way to verify a quantum computer is through formal verification, running algorithms \cite{shor, jozsa, grover, grover_2} with known outcomes for a known input. However, running a single algorithm and getting a satisfactory outcome is not a particularly rigorous test of the full function of a quantum computer.  What is required are classes of formal verification tests with effectively infinite variability where the output is known for a given input.

In this work we suggest a class of verification tests that are based on a quantum evolution by a finite set of quantum gate operations.
An arbitrary sequence of these operations are applied to a quantum state-vector, which is initially fully contained within an invariant Hilbert space of the set of gate operations.
Then the ``leak'' of the state out of the invariant subspace is measured, and used to assess the fidelity of the corresponding set of quantum gate operations.
Here, we consider three such sets of quantum gate operations generated by all possible $CP$ (controlled-phase), $CNOT$ (controlled-NOT) and the $SWAP^\alpha$ (power-of-SWAP) quantum gate operations on an $n$-qubit system, respectively.
These $2$-qubit quantum gates are comonly used in basic-gate sets \cite{set_0,set_1, set_2} for universal, gate-based \cite{uni1,uni2,uni3,uni4} quantum computing. Measuring their performance is critical for verifying the operation of NISQ \cite{NISQ, NISQ_2, google_supreme} and early fault-tolerant quantum computers \cite{verify_DFS1, fault_tolerant, set_2}.

In this work, we begin by identifying the groups formed by each of the three $2$-qubit quantum gate operations, mentioned above. We then determine the invariant Hilbert subspaces corresponding to each of the three groups.
For the $CP$ operation, we determine $2^n$ $1$-dimensional invariant subspaces corresponding to the computational state-vectors. For the $CNOT$ operation, we determine two $2$-dimensional invariant subspaces and one $(2^n-2)$-dimensional invariant subspace. For the $SWAP^\alpha$ operation we find a number of $O(n^2)$ distinct invariant subspaces, and propose a recursive algorithm to construct explicitly these subspaces.
Then we use these invariant subspaces to outline a verification procedure for quantum logic hardware.

The paper is organised as follows:
In the next Sec. \ref{sec:theory}, we define our approach in associating a set of quantum gate operations with a group of actions.
In Sec. \ref{sec:results} we present our analysis of the group theoretic properties of the $CP$ (Sec. \ref{sec:CP}), the $CNOT$ (Sec. \ref{sec:CNOT}) and the $SWAP^\alpha$ (Sec. \ref{sec:SWAP}) gate operations, and outline our verification procedure (Sec. \ref{sec:verify}).
We present our concluding remarks in Sec.\ref{sec:conclusion}.

\section{Theoretical approach and notation}\label{sec:theory}

A quantum gate operation is a unitary map on the Hilbert space of a qubit system. Given a set $\mathcal{S}$ of quantum gate operations, there is an associated group of unitary maps generated by the elements of $\mathcal{S}$: that is, the group of all the maps which can be formed by sequentially performing a finite number of operations in $\mathcal{S}$ as well as their inverses.
For an $n$-qubit system, we will denote the groups associated with the sets of  $CP$, $CNOT$ and $SWAP^\alpha$ gate operations by $CP^{(n)}$, $CNOT^{(n)}$ and $SWAP^{\alpha(n)}$, respectively.

To determine the elements and orders of these groups we must find all unique operations that can be performed with the corresponding quantum gate operations. Since in general we will be concerned with non-commutative $2$-qubit gate operations, the group elements will consist of the gate operations over all possible \textit{ordered} pairs of qubits, together with all unique combinations of these operations.
We will denote gate operations over a pair of qubits $i$ and $j$, as $CP^{(n)}_{ij}$, $CNOT^{(n)}_{ij}$ and $SWAP^{\alpha (n)}_{ij}$.
For a $2$-qubit system the $CNOT^{(2)}$ group will consist of the two $CNOT$ operations, $ CNOT_{0,1}^{(2)}$ and $ CNOT_{1,0}^{(2)}$, and their unique distinct combinations, $ CNOT_{0,1}^{(2)} \times CNOT_{1,0}^{(2)}$, $ CNOT_{1,0}^{(2)} \times CNOT_{0m1}^{(2)}$ and $CNOT_{0,1}^{(2)} \times CNOT_{1,0}^{(2)} \times CNOT_{0,1}^{(2)}$.

Throughout this work we will work with the ``natural'' matrix representations of the $CP^{(n)}$, $CNOT^{(n)}$ and $SWAP^{\alpha(n)}$ groups.
These representations are the $2^n \times 2^n$ matrix representations whose elements act on the $2^n$ dimensional state-vector, representing the quantum state of a $n-$ qubit system, with state-vector terms corresponding to the $2^n$ computational basis states.

\section{Results}\label{sec:results}

\subsection{The $CP^{(n)}$ group and invariant subspaces}\label{sec:CP}

The $CP$ (controlled-phase) gate is a $2$-qubit quantum gate that performs a controlled $z$-rotation by $\pi \ rad$ on a target qubit if a control qubit is in the state $|1\rangle$.
The $CP$ is a maximally entangling gate, capable of transforming separable states into maximally entangled states.
Therefore it is extensively used as an entagling gate in basic-gate sets \cite{Rigetti} for universal gate-based quantum computation,
and in measurement-based quantum computation \cite{measure_1, measure_2, fault_tolerant} to construct partially entangled cluster states\cite{measure_1}.

The $CP$ operations are invariant under exchange of the control and the target qubits, and are their own inverses. This means that the $CP^{(2)}$ group has only one generator of order $2$. Hence the $CP^{(2)}$ group is isomorphic to the cyclic group of order $2$, which is denoted by $C_2$.
The $CP^{(n)}$ group is generated by the $n(n-1)/2$ distinct $CP$ operations on $n$-qubits, which are all group elements of order $2$. Since these operations commute, $CP^{(n)}$ is an abelian group. Moreover, these operations form a minimal generating set: that is, none of the operations can be written as a product of the others and their inverses. Then, given that each $CP$ operation has order $2$, it follows that the $CP^{(n)}$ group is isomorphic to the direct product of $n(n-1)/2$ cyclic groups of order $2$: $CP^{(n)}\cong C_2^{n(n-1)/2}$. The order of the $CP^{(n)}$ group is given by
\begin{equation}\label{eq:CP_order}
|CP^{(n)}| = 2^{n(n-1)/2}
\end{equation}

The matrices in the matrix representation of the $CP^{(2)}$ group are
\begin{equation}
CP^{(2)}_{0,1} \equiv \Bigg(
    \begin{smallmatrix}
		1 & 0 & 0 & 0 \\
		0 & 1 & 0 & 0 \\
		0 & 0 & 1 & 0 \\
		0 & 0 & 0 & -1 \\
\end{smallmatrix} \Bigg) \mathrm{, and }\  {CP^{(2)}_{0,1}}^2=\Bigg(
\begin{smallmatrix}
         1 & 0 & 0 & 0 \\
         0 & 1 & 0 & 0 \\
         0 & 0 & 1 & 0 \\
         0 & 0 & 0 & 1 \\
\end{smallmatrix}\Bigg).
\end{equation}
Similarly the matrix representation of the $CP^{(n)}$ group, for $n>2$, also contains only diagonal matrices with $\{-1, +1\}$ entries.
Therefore each computational basis state-vector spans an $1$-dimensional invariant Hilbert subspace by itself.

\subsection{The $CNOT^{(n)}$ group and invariant subspaces}\label{sec:CNOT}

The $CNOT$ operation is a $2$-qubit quantum gate which flips the state of a target qubit if a control qubit is in the state $|1\rangle$.
In the computational basis, the two generating elements of $CNOT^{(2)}$ are represented by the following matrices:
\begin{equation}
CNOT^{(2)}_{1,0}=\Bigg(
\begin{smallmatrix}
         1 & 0 & 0 & 0 \\
         0 & 1 & 0 & 0 \\
         0 & 0 & 0 & 1 \\
         0 & 0 & 1 & 0 \\
\end{smallmatrix}\Bigg), \
        CNOT^{(2)}_{0,1}=\Bigg(
\begin{smallmatrix}
         1 & 0 & 0 & 0 \\
         0 & 0 & 0 & 1 \\
         0 & 0 & 1 & 0 \\
         0 & 1 & 0 & 0 \\
\end{smallmatrix}\Bigg).
\end{equation}
Like the $CP$ gate, the $CNOT$ gate is maximally entangling, capable of transforming separable states to maximally entangled states. It is perhaps the most commonly implemented $2$-qubit gate in gate-based quantum computers \cite{set_0, set_1, set_2, IBM}, since its operation as a controlled-NOT is logically intuitive, and convenient for designing quantum circuits

In order to investigate the $CNOT^{(n)}$ group, it is useful to associate each computational basis state-vector with an element of $\mathbb{F}_2^n$, the $n$-dimensional vector space over the field with $2$ elements. We do this in the natural way: for example, we associate the state-vector $\ket{010}$ with the vector $(0,1,0)$. Since each $CNOT$ operation sends the computational basis to itself (each basis state-vector is transformed to a basis state-vector), we can further associate each element $g\in CNOT_n$ with a corresponding function, call it $\theta(g)$, on $\mathbb{F}_2^n$. It can be shown (see appendix \ref{app:CNOT_iso}) that $\theta(g)\in GL(n,2)$, the group of invertible linear maps from $\mathbb{F}_2^n$ to itself, and moreover that the map $\theta:CNOT^{(n)}\to GL(n,2)$ is a group isomorphism. Hence $CNOT^{(n)}\cong GL(n,2)$.

By inspection we find that $CNOT^{(n)}$  has two one-dimensional invariant subspaces:  $V_0 = \text{span}\big\{|0\rangle\big\}$ and $V_1 = \text{span}\big\{\frac{1}{\sqrt{2^n-1}}\sum_{i=1}^{2^n-1} |i\rangle\big\}$. The invariance of $V_0$ is evident, while for $V_1$ one should note that each $CNOT$ operation is a bijection (one-to-one and onto) between all computational basis states, except the zeroth state.
Furthermore it can be shown (see appendix \ref{app:CNOT_irrep}) that the Hilbert space orthogonal to $V_0$ can be decomposed into two irreducible invariant subspaces, one of which is $V_1$. Therefore we deduce that the $(2^n-2)$-dimensional subspace, $V_2$, that is orthogonal to $V_0$ and $V_1$, is itself an irreducible invariant subspace.
Hence the action of $CNOT^{(n)}$ on the Hilbert space of $n$ qubits has three irreducible invariant subspaces that can be defined in terms of basis vectors as
\begin{equation}
V_0 = \text{span} \big\{ |0\rangle \big\}
\end{equation}
\begin{equation}
V_1=\text{span} \big\{|v_1\rangle \big\} \text{, where } |v_1\rangle=\frac{1}{\sqrt{2^n-1}}\sum_{i=1}^{2^n-1} |i\rangle
\end{equation}
\begin{equation}
V_2=\text{span} \Bigg\{\frac{\sqrt{2^n-1}|i\rangle - |v_1\rangle}{2^{n/2}}:i=1,...,2^n-1  \Bigg\}
\end{equation}
We can also use the isomorphism of $CNOT^{(n)}$ to $GL(n,2)$ to find the order of the $CNOT$ group. For large numbers of qubits, $n$, it can approximated as
\begin{equation} \label{orderGLn2}
	|CNOT^{(n)}|= |GL(n,2)|=\prod_{i=0}^{n-1} (2^n-2^i) \approx 0.29 \times 2^{n^2}
\end{equation}

\subsection{The $SWAP^{\alpha (n)}$ group and invariant subspaces}\label{sec:SWAP}

The $SWAP^\alpha$ is a $2$-qubit quantum gate operation that continuously exchanges the values of two qubits as $\alpha$ is varied.
The action of the $SWAP^\alpha$ on a $2$-qubit system can be illustrated by its matrix representation:
\begin{equation}\label{eq:swap_alpha}
SWAP^{\alpha (2)}_{01} \equiv \Bigg(
\begin{smallmatrix}
     1 & 0 & 0 & 0 \\
     0 & \frac{1}{2} \left(1+e^{i \pi  \alpha }\right) & \frac{1}{2} \left(1-e^{i \pi  \alpha }\right) & 0 \\
     0 & \frac{1}{2} \left(1-e^{i \pi  \alpha }\right) & \frac{1}{2} \left(1+e^{i \pi  \alpha }\right) & 0 \\
     0 & 0 & 0 & 1 \\
\end{smallmatrix} \Bigg).
\end{equation}
The $SWAP^\alpha$ gate has non-zero entangling power for non integer values of $\alpha$, so it can be used as an entangling gate in basic-gate sets for universal gate-based quantum computing.
It is often implemented in spin-qubit quantum computing architectures \cite{spin_qubit_1, spin_qubit_2, spin_qubit_3, spin_qubit_4}, since it arises naturally from the spin exchange interaction \cite{exchange_int_1, exchange_int_2, exchange_int_3, exchange_int_4}. Finding a group isomorphism and the invariant subspaces for the $SWAP^{\alpha (n)}$ group is challenging for a general value of $\alpha$. Therefore we consider the simplest case of $\alpha=1$.

\subsubsection{The $SWAP^{(n)}$ group and invariant subspaces}

The $SWAP$ is a $2$-qubit quantum gate operation which completely exchanges the values of two qubits, and has zero entangling power.
The action of the $SWAP^{(n)}$ group on a $n$-qubit system is isomorphic to $S_n$, the group of permutations over $n$ distinguishable objects (this is straightforward to see by regarding each qubit as a distinguishable object).
To determine the invariant subspace structure of $SWAP^{(n)}$, we first note that the $SWAP$ operation conserves the Hamming weight (the number of qubits in state $|1\rangle$) of a state.
Therefore, all states with Hamming weight $i$ span an invariant subspace, $V_i$, of order
\begin{equation} \label{orderVi}
|V_i|=\frac{n!}{(n-i)!i!}=\binom{n}{i}.
\end{equation}
However $V_i$ can be further decomposed to smaller, irreducible, invariant subspaces.
Using the fact that $SWAP^{(n)}\cong S_n$, we show, in appendix \ref{sec:V_i}, that for $i \leq \lfloor \frac{n}{2} \rfloor $, each $V_i$ can be decomposed as
\begin{equation} \label{keyR}
V_i=V_{i,0}\oplus V_{i,1}\oplus..V_{i,i},
\end{equation}
where $V_{i,j}$ are irreducible invariant subspaces, and subspaces with the same second subscript, $j$, correspond to the same irreducible representation (irrep) of $SWAP^{(n)}$. This implies that
\begin{equation}\label{eq:keyR.0.1}
|V_{i,j}|=|V_{i',j}| \text{ for any }j \leq i <i'	.
\end{equation}
For $i \geq \ceil{\frac{n}{2}} $, the irreducible invariant subspaces $V_{i,j}$ are identical upon flipping the values of all qubits. Therefore our analysis will consider only the case $i \leq \lfloor \frac{n}{2} \rfloor $.
From Eq. (\ref{keyR}), it follows that the total number of 
irreducible invariant subspaces is
\begin{equation} \label{eq:Ninv}
N=\begin{cases}
\sum_{i=0}^{\frac{n}{2}-1}(i+1)+\frac{n+2}{2}=\frac{(n+2)^2}{4} \textrm{, n even}
\\
\\
2\sum_{i=0}^{\frac{n-1}{2}}(i+1)=\frac{(n+1)(n+3)}{4} \textrm{, n odd,}
\end{cases}
\end{equation}
and that the number of irreducible invariant subspaces $V_{ij}$ for a given value of $j$ is 
\begin{equation}\label{eq:Nvij}
N_j = |n-2j|+1.
\end{equation}
From Eq. (\ref{eq:keyR.0.1}) it follows that the dimensions of the $V_{ij}s$ are given by
\begin{equation}\label{eq:size_vij}
|V_{i,j}| = \begin{cases}
\binom{n}{j} \text{ , for }j=0\\
\binom{n}{j} - \binom{n}{j-1} \text{ , for }1\leq j\leq n/2.
\end{cases}
\end{equation}

Based on Eqs. (\ref{keyR}) and (\ref{eq:keyR.0.1}), and using that the subspaces $V_{i,j}$ and $V_{i',j}$ correspond to the same irreducible representation of $SWAP^{(n)}$, we designed and implemented a recursive computational procedure, outlined in appendix \ref{sec:procedure}, to find explicit sets of basis vectors for each of the $V_{i,j}s$.

\begin{figure}[h]
    \centering
    \includegraphics[width=9cm]{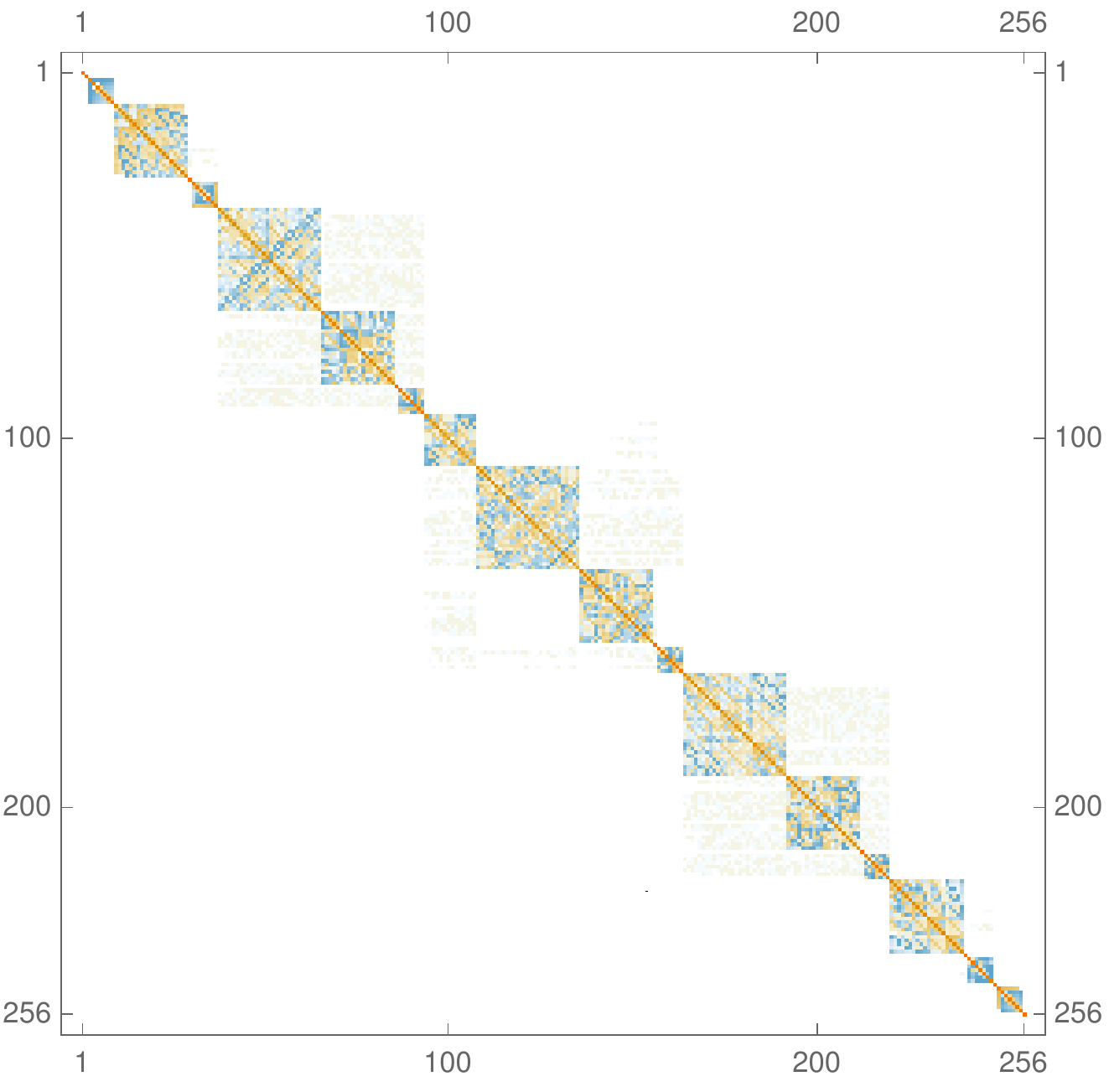}
    \caption{Matrix color plot of the $2^8 \times 2^8$ block diagonalized matrix representation of the $SWAP^{(8)}$ group.
The matrix plot is obtained by summing and block-diagonalizing large number of matrices from the matrix representation of the $SWAP^{(8)}$.
Blue-green elements correspond to negative values. Yellow-red elements correspond to positive values.
Pale coloured matrix elements outside the diagonal blocks, correspond to small value rounding errors.}
\label{fig:S8cm}
\end{figure}

We demonstrate our procedure with the example of the $SWAP^{(8)}$ group. We find bases for its $V_{ij}$s, and use these bases to construct a transformation matrix, which we use to block-diagonalize the matrix representation of the $SWAP^{(8)}$ group.
The transformed block-diagonal form of the matrix representation of $SWAP^{(8)}$ is given in the form of a matrix plot in Fig. \ref{fig:S8cm}.

Each diagonal block in the transformed matrix in Fig. \ref{fig:S8cm} corresponds to an irreducible invariant subspace $V_{ij} $ (ordered, from left to right, in terms of increasing $i$, and decreasing $j$).
Therefore the number of occurrences and the dimensions of the blocks should match those of the $V_{ij}$s, given by Eqs. \eqref{eq:Nvij} and \eqref{eq:size_vij}, respectively.
It can be verified by inspection that this is indeed true.

\subsubsection{Invariant subspaces of $SWAP^{\alpha }$}
It can be shown that the $SWAP^{\alpha (n)}$ has the same irreducible invariant subspaces for any real $\alpha \neq 0$, including the case of $\alpha=1$.
To see why this is true, we decompose the matrix representation of $SWAP^{\alpha (2)}_{01}$, given in Eq. \eqref{eq:swap_alpha}, as
\begin{equation}
SWAP^{\alpha (2)}_{01} = a \times SWAP^{(2)}_{01} + b \times I^{(2)}
\end{equation}
where $a=\frac{1}{2}(1+e^{i \pi \alpha})$, $b=\frac{1}{2}(1-e^{i \pi \alpha})$ and $I$ is the identity. This decomposition is true for any number of qubits, $n$, so we can write
\begin{equation}\label{eq:decompose_2}
SWAP^{\alpha (n)}_{pq} = a \times SWAP^{(n)}_{pq} + b \times I^{(n)}
\end{equation}
Now, suppose $\vec{u}$ is a state-vector lying in an irreducible invariant subspace $V_{i,j}$ of $SWAP^{(n)}$. Then
\begin{align}
SWAP^{\alpha (n)}_{pq}\vec{u} &= a \times SWAP^{(n)}_{pq} \vec{u} + b \vec{u} \\
&\in V_{i,j}
\end{align}
since the invariance of $V_{i,j}$ under $SWAP^{(n)}$ implies that $SWAP^{(n)}_{pq} \vec{u}\in V_{i,j}$. Hence the invariant subspaces of $SWAP^{\alpha (n)}$ are contained in those of $SWAP^{(n)}$. Conversely, provided $\alpha\neq 0$, so that $a\neq 0$, we may invert (\ref{eq:decompose_2}) to get
\begin{equation}
SWAP^{(n)}_{pq} = \frac{ SWAP^{\alpha (n)}_{pq} - b \times I^{(n)}}{a},
\end{equation}
and the previous argument shows that the invariant subspaces of $SWAP^{(n)}$ are contained in those of $SWAP^{\alpha(n)}$.
Therefore $SWAP^{(n)}$ and $SWAP^{\alpha(n)}$ share the same irreducible invariant subspaces $V_{i,j}$.

\subsection{Invariant subspace verification test}\label{sec:verify}

In this section we outline a procedure that uses our knowledge of the invariant subspaces of a particular quantum gate operation (e.g. CNOT ) to verify the performance of a quantum computer.

This procedure consists of the following $3$ steps:
\begin{enumerate}

\item Initialize the quantum computer in a state that is fully confined within one (or a few) of the irreducible invariant subspaces of the operation.

\item Perform randomly (between different pairs of qubits) multiple operations. The greater the number of operations the more rigorous the test.

\item Perform a measurement that projects the quantum computer state onto the basis states of the irreducible invariant subspaces of the operation.
\end{enumerate}

If the gate operations are implemented perfectly, then the state of the quantum computer should remain confined within the initial irreducible invariant subspaces. However, in practice the gate operations would be implemented with fidelity less than one. Therefore, after multiple operations, the state of the quantum computer will ``leak'' out of the initial irreducible invariant subspaces, and will have non-zero projection in the rest of the Hilbert space. This  projection is determined by a measurement in the bases of the irreducible invariant subspaces, and can be used as a measure for the fidelity of the operation.

We note that on the current NISQ computers, the initialization and the measurement steps, $1$ and $3$, respectively, might incur an error of a comparable magnitude to the error incurred from the multiple gate operations.
A possible solution to this problem would be to use POVMs \cite{POVM_1, POVM_2, POVM_3, POVM_4} followed by post-processing, to initialize and measure the state in steps 1 and 3, respectively.

\subsubsection{Verification with CP}

As noted in Sec.\ref{sec:CP} the individual $n$-qubit computational basis states are $1$-dimensional invariant subspaces under the action of the $CP^{(n)}$ group. This means that multiple $CP$ operations do not change the $Z$-basis measurement probabilities. Therefore, the verification procedure outlined above will require simply (1) measurement in the $Z$-basis, (2) application of multiple randomly chosen $CP^{(n)}$ operations, (3) measurement in the $Z$-basis. Any deviation from the measurement probabilities will indicate an error.
Since the $CP$ operation can be created in a number of different ways, for example from a combination of CNOT operations and single-qubit operations, this simple test can be used to test multiple operations of a quantum computer.

\subsubsection{Verification with CNOT}
As shown in Sec. \ref{sec:CNOT} the $CNOT^{(n)}$ group has a large $(2^n -2)$-dimensional irreducible invariant subspace. This implies that the $CNOT$ operation alone is of limited value in our verification procedure described above. Even imperfect $CNOT$ operations acting on a qubit state, initialized within the large subspace, would be likely to produce small projections onto the two $1$-dimensional invariant subspaces. Alternatively initializing a state in either of the two $1$-dimensional invariant subspaces would be a useful test, but not as comprehensive as the $CP$ operation.

\subsubsection{Verification with $SWAP^\alpha$}
The case of the $SWAP^\alpha$ is the most interesting and resourceful when it comes to invariant subspaces and their use in our verification procedure. The most simple procedure involving the $SWAP^\alpha$ would be to check if multiple applications of randomly chosen operations conserves the Hamming weight of the initial state. This would correspond to testing the invariance of the $V_i$ subspaces.
A more complicated and comprehensive test would utilize the irreducible invariant subspaces $V_{ij}$. Such a test would require a more elaborate procedure to initialize the state in a given irreducible invariant subspace $V_{ij}$ and subsequently to perform a measurement projecting onto the basis of this subspace.
Again, this test can be made more comprehensive by constructing the $SWAP^\alpha$ operation from combinations of the other entangling gates and single-qubit operations.

\section{Conclusion}\label{sec:conclusion}

In this work we analysed the operation of the $CP$, the $CNOT$ and the $SWAP^\alpha$ quantum gate operations from a group theoretic point of view.
We found that the group of $CP$ operations on $n$-qubits is isomorphic to the direct product of  $n(n-1)/2$ cyclic groups of order $2$. We determined that its irreducible invariant subspaces correspond to the individual computational basis states-vectors.
We found that the group of $CNOT$ operations on $n$-qubits is isomorphic to the general linear group of $n$-dimensional space over a field with two elements, $GL(n,2)$. We used this result to demonstrate that the group generated by $CNOT$ operations on $n$ qubits has one $(2^n-2)$-dimensional and two $1$-dimensional irreducible invariant subspaces.
For the $SWAP^\alpha$ operation we showed that its irreducible invariant subspaces are the same for all values of $\alpha$.
We therefore investigated the simpler case of the $SWAP$ operation and constructed a method to determine its irreducible invariant subspaces.

For each group we considered, we suggested how to construct verification tests for the operation of a quantum  computer, using the invariant subspaces discovered. 
These tests initialize a state in a particular invariant subspace, and measure by how much the invariant subspace is violated by multiple applications of the corresponding quantum gate operations.
We believe that these tests will be important for verifying the operation of large NISQ and early fault-tolerant quantum computers.

\begin{acknowledgements}
YY acknowledges financial support from the Hitachi Cambridge Laboratory through an iCASE studentship RG97399 (voucher 18000078), and the Engineering and Physics Research Council (EPSRC). CHWB and YY would like to thank Dr Ross Lawther for his guidance in constructing the invariant subspaces for both the $CNOT$ and the $SWAP^{\alpha}$ operations. We also would like to thank A. Lasek, D. Arvidsson-Shukur, H. Lepage and N. Devlin for useful discussions.
\end{acknowledgements}

\appendix

\section{Proof that $CNOT^{(n)}\cong GL(n,2)$}\label{app:CNOT_iso}

For each $g\in CNOT^{(n)}$, let $\theta(g)$ be the function from $\mathbb{F}_2^n\to\mathbb{F}_2^n$ obtained by associating computational basis elements with elements of $\mathbb{F}_2^n$ as previously described. First note that for all $f,g\in CNOT^{(n)}$, we have $\theta(f\circ g)=\theta(f)\circ \theta(g)$ (this follows trivially from the one-to-one association between the computational basis and $\mathbb{F}_2^n$). Hence, if we can show that $\theta(CNOT^{(n)}_{ij})\in GL(n,2)$ for all $i,j$, then since the $CNOT^{(n)}_{ij}$s generate $CNOT^{(n)}$ it will follow that $\theta(g)\in GL(n,2)$ for all $g\in CNOT^{(n)}$.\\
\\
Let $g=CNOT_{ij}^{(n)}$. Since $\theta(g)$ leaves all but the $i^{th}$ and $j^{th}$ entries unaffected, it suffices to consider only the $2$-qubit case with $g=CNOT^{(2)}_{01}$, and show that $\theta(g)$ is linear and invertible. To do so, we simply write down the effect of $\theta(g)$ on each element of $\mathbb{F}_2^2$:  $(0,0)\mapsto(0,0)$, $(0,1)\mapsto(0,1)$, $(1,0)\mapsto(1,1)$ and $(1,1)\mapsto(1,0)$. One can easily see that $\theta(g)$ is invertible, and remembering that addition is modulo $2$, $\theta(g)$ is also linear as required. \\
\\
So $\theta$ maps into $GL(n,2)$, and since it is structure-preserving (i.e. $\theta(f\circ g)=\theta(f)\circ \theta(g)$) it is a group homomorphism from $CNOT^{(n)}\to GL(n,2)$. In order to show that $\theta$ is an isomorphism, we must further show that it is a bijection. Injectivity is immediate, since $\ker \theta=\{\text{id}\}$. In order to show surjectivity, it suffices to show that im$\theta$ contains a generating set. It can be shown that $GL(n,2)$ is generated by the linear maps $m_1$ and $m_2$ given in the standard basis by the matrices
$M_1:= \Bigg( \begin{smallmatrix} 1 &  &  &  &  \\
				     & 1 &  &  &  \\
				      &  &  & \ddots & \\
				    1 &  &  &  & 1
\end{smallmatrix}\Bigg)$
and
$M_2:=\Bigg(\begin{smallmatrix} 0 & 1 &  &  &  \\
& 0 & 1 &  &  & \\
&  &  & \ddots & 1 \\
1 &  &  &  & 0
\end{smallmatrix}\Bigg). $

Since $m_1=\theta(CNOT^{(n)}_{n-1\ 0})$, $m_1\in$im$\theta$. The map $m_2$ acts on elements of $\mathbb{F}_2^n$ by applying the permutation (01...n-1) to entries. Since $CNOT^{(n)}_{ij}CNOT^{(n)}_{ji}CNOT^{(n)}_{ij}=SWAP^{(n)}_{ij}$, the group $CNOT^{(n)}$ contains all $SWAP$s and hence im$\theta$ contains all maps which are transpositions of tuple entries. Since transpositions generate $S_n$, we conclude that $m_2\in$im$\theta$ and hence that $\theta$ is surjective, finishing the proof.
%%%%%%%%%%%%%%%%%%%%%%%%%%%%%%%%%%%%%%%%%%%%%%%%

\section{Irreducible invariant subspaces of $CNOT^{(n)}$}\label{app:CNOT_irrep}

Here we consider the decomposition of the Hilbert space of $n$-qubits to subspaces that are invariant under the action of the $CNOT^{(n)}$ group. First we note that the $CNOT$ operations do not affect the zeroth state $|0\rangle =|0 0 .. 0\rangle$, so it spans a 1-dimensional invariant subspace $V_0 =span\{|0\rangle\}$, on its own. Let us denote the set of computational basis states excluding the zeroth as $X$, so that $X = \big\{ |i\rangle : i = {1, ...,2^n-1}\big\}$, and the Hilbert space spanned by the set as $V_0^\bot$.
To find the decomposition to irreducible invariant subspaces of $V_0^\bot$, we first show that the action of the $CNOT^{(n)}$ group on $X$ is doubly-transitive \footnote{An action of a group on a set of elements is doubly-transitive if for any two ordered tuples, each having a pair of distinct elements from the set, there is a group element taking one ordered tuple to the other}:\\
\\
\textit{Proof.} Note that it suffices to provide a single tuple of states $\big(|\psi_1\rangle, |\psi_2\rangle\big)$ such that any other tuple $\big(|\psi_1'\rangle, |\psi_2'\rangle\big)$ with $|\psi_1'\rangle \neq |\psi_2'\rangle$ may be obtained by successive application of $CNOT$ gates: double-transitivity will then follow. Consider $\big(|\psi_1\rangle, |\psi_2\rangle\big) = \big(|010..0\rangle, |100..0\rangle)$. Using $CNOT$ operations with the zeroth and first qubits as control qubits, we can change the values of the other $n-2$ qubits of each state separately, and take the initial tuple to any other tuple of states where the first two qubits are unchanged. Therefore we only need to show that $CNOT^{(2)}$ acts doubly-transitively on the set $\{|01\rangle, |10\rangle, |11\rangle\}$. This can be verified easily by hand, completing the proof.\\
\\
We now use proposition 4.4.4 from \cite{representation_symmetries}, which states that for a group $G$ that acts doubly-transitively on a set of vectors $S$, the space spanned by $S$ decomposes to two irreducible invariant subspaces. Transferring this result to the context of our problem, it means that $V_0^\bot$ decomposes to two irreducible invariant subspaces under the action of $CNOT^{(n)}$.  \\
\\
Finally we note that the state vector $v_1= \frac{1}{\sqrt{2^n-1}}\sum_{i=1}^{2^n-1} |i\rangle$ is invariant under $CNOT^{(n)}$ because each $CNOT$ operation is a bijection (one-to-one and onto) between all computational basis state-vectors, except the zeroth state-vector. Therefore the $(2^n-2)$-dimensional subspace, $V_2$, that is orthogonal to both $V_0$ and $V_1$, is an irreducible invariant subspace.

\section{Irreducible invariant subspaces of $SWAP^{(n)}$}\label{sec:V_i}

Since $SWAP$ operations conserve the Hamming weight of quantum states, the subspace $V_i$ spanned by all state vectors of Hamming weight $i$ is invariant under $SWAP^{(n)}$. However $V_i$ can be decomposed further to smaller invariant subspaces.\\
\\
Consider the action of the group $SWAP^{(n)}$ on $n$ qubits.
For \(i \leq \frac{n}{2}\), let \(x_i\) be the set of $i$-element subsets of \(X\) (so that the action of $S_n$ on \(x_i\) is isomorphic to the action of $SWAP^{(n)}$ on $V_i$).\\
\\
Let \(\pi_i\) be the permutation representation character of the action of \(S_n\) on \(x_i\).
The Hermitian product of two such characters $\pi_k$ and $\pi_l$ is given by
\begin{equation} \label{H.pr.r}
\langle\pi_k,\pi_l \rangle=\frac{1}{|S_n|}\sum_{s \in S_n} \pi_k(s)\pi_l(s) = \langle \pi_k \pi_l, 1_G \rangle =l+1
\end{equation}
where \(0\leq l \leq k \leq \frac{n}{2}\), and \(1_G\) denotes the trivial representation.\\
\\
Fix \(k\leq \lfloor{n} \rfloor\) and assume for our inductive hypothesis that for \(0\leq i \leq k-1\),
\begin{equation} \label{induction}
    \pi_i=\chi^{(n,0)}+\chi^{(n-1,1)}+...+\chi^{(n-i,i)}
\end{equation}
where the \(\chi \)s are irreducible characters (characters of irreducible representation of $S_n$).\\
\\
For $r=0$, \(x_0\) has one element so \(S_n\) acts trivially on it, thus $\pi_0=1_G$. This implies that $\chi^{(n,0)}=1_G$.\\
\\
For \(1\leq i \leq k-1\), writing \(\chi^{(n-i,i)}=\pi_i-\pi_{i-1}\), and using (\ref{H.pr.r}) we get that
\begin{equation}
\langle \pi_k,\chi^{(n-i,i)} \rangle= \langle\pi_k,\pi_i \rangle-\langle \pi_k,\pi_{i-1} \rangle=1 .
\end{equation}
Therefore \(\chi^{(n-i,i)}\) is a component of \(\pi_k\) with multiplicity \(1\). Hence we can write
\begin{equation}\label{eq:B4}
    \pi_k=\chi^{(n,0)}+\chi^{(n-1,1)}+...+\chi^{(n+1-k,k-1)}+\chi'
\end{equation}
for some \(\chi'\).\\
\\
But $ \langle \pi_k,\pi_k \rangle =k+1$ from (\ref{H.pr.r}), and $\langle \pi_k,\pi_k \rangle=k+\langle \chi',\chi' \rangle$ from (\ref{eq:B4}), so $ \langle \chi',\chi' \rangle=1$. Therefore \(\chi'\) is an irreducible character which we denote as \(\chi^{(n-k,k)}\). Hence:
\begin{equation}
    \pi_k=\chi^{(n,0)}+\chi^{(n-1,1)}+...+\chi^{(n-k,k)}
\end{equation}
where each \(\chi\) is an irreducible character (corresponding to an irreducible invariant subspace). Thus the inductive step is complete. This result implies that for an $n$-qubit system, $V_i$ decomposes into irreducible invariant subspaces, under $SWAP^{(n)}$, as
\begin{equation}
V_i=V_{i,0}\oplus V_{i,1}\oplus..V_{i,i},
\end{equation}
where subspace $V_{i,j}$ corresponds to irrep $\chi^{(n-j,j)}$.

\section{Constructing basis state vectors for the irreducible invariant subspaces of $SWAP(n)$} \label{sec:procedure}

The Hilbert subspaces $V_i$ corresponding to $n$ qubit states of Hamming weight $i$ are invariant under the action of $SWAP^{(n)}$. However, as proved in appendix \ref{sec:V_i}, the subspaces $V_i$ can be decomposed further as $
V_i=V_{i,0}\bigoplus V_{i,1}\bigoplus...\bigoplus V_{i,i}
$
where \(V_{i,j}\) are irreducible invariant subspaces, and the second subscript, $j$, denotes correspondence to the same irrep. of $SWAP^{(n)}$. In particular, we have $
|V_{i,j}|=|V_{i',j}| \text{ for any }j \leq i <i'
$.
Below we outline a procedure to construct a set of basis state vectors for the subspaces $V_{i,j}$ for an $n-$qubit system. We consider the case of $i \leq \lfloor \frac{n}{2} \rfloor $ only, since the case for $i > \lfloor \frac{n}{2} \rfloor $ is identical upon global qubit flip.
\subparagraph{\textbf{Constructing basis state vectors for $V_{i,j}$}}

\begin{enumerate}[ labelwidth=!, labelindent=0pt]

\item For $i=0$ , we have the $1$-dimensional invariant subspace $V_0$ spanned by the zeroth state-vector
\begin{equation}
V_0 = V_{0,0} = span\{|0..0\rangle\}
\end{equation}
\item For $i=1$, \(|V_1|=n\), and \(V_1=V_{1,0}\bigoplus V_{1,1}\). Also $|V_{0,0}|=|V_{1,0}|=1$ and $|V_{1,1}|=|V_1|-|V_{0,0}|=n-1$. The single state vector of \(V_{1,0}\) can be written as the sum of all computational state-vectors in $V_1$ (all state-vectors with Hamming weight $1$)
\begin{equation}
V_{1,0} =span\Big\{ \frac{1}{\sqrt{n}}\sum_{|\phi \rangle \in V_1} |\phi\rangle \Big\} = \frac{1}{\sqrt{n}}\sum_{i=0}^{n-1} |..0_{i-1}1_i 0_{i+1}..\rangle \Big\}
\end{equation}
\(V_{1,1}\) can be determined by taking an arbitrary set of basis state vectors for the orthogonal compliment of \(V_{1,0}\) in \(V_1\).\\
\\
\item For $i\geq 2$,  $V_i=V_{i,0}\bigoplus V_{i,1}\bigoplus...\bigoplus V_{i,i}$ and $|V_i|=\binom{n}{i}$.
Let $V_{i,i}^\bot$ denote the orthogonal complement of $V_{i,i}$ in \(V_i\).\\
\\
First we need to find sets of basis state vectors that span $V_{i,i}$ and $V_{i,i}^\bot$. Note that $|V_{i,i}^\bot|=|V_{i-1}|$, since the two spaces consist of irreducible subspaces that correspond to the same irreps of $SWAP^{(n)}$ ($ V_{i,i}^\bot=V_{i,0} \bigoplus V_{i,1}\bigoplus ...\bigoplus V_{i,i-1}$ and $ V_{i-1} = V_{i-1,0} \bigoplus V_{i-1,1} \bigoplus ...\bigoplus V_{i-1,i-1}$, respectively).
Furthermore, this means that we can construct basis state-vectors for $V_{i,i}^\bot$ such that they transform, under $SWAP$ operations, in the same way as the computational state-vectors in $V_{i-1}$ (the state-vectors with Hamming weight $i-1$).
Then we will be able to decompose $V_{i,i}^\bot$ in the same way as we decomposed $V_{i-1}$. In practice this can be conveniently implemented recursively. \\
\\

The basis state-vectors for $V_{i,i}^\bot$ can be constructed in the following way:

\begin{enumerate}

\item Denote the $\binom{n}{i-1}$ basis state-vectors for $V_{i,i}^\bot$ by $v^i_{s_k}$, where $\{s_k\}$ are all subsets of  size $i-1$ of the set $\{0,..,n-1\}$, for $ k = 0,...,\binom{n}{i-1}-1$; e.g. for $n=4, i=2$: \(s_0=\{0\}\), \(s_1=\{1\}\), \(s_2=\{2\}\),  \(s_3=\{3\}\).

\item Construct $v^{(i)}_{s_k}$ by summing over all computational state-vectors, with Hamming weight $i$, whose qubits in positions given by the elements of $s_k$ are in the $|1 \rangle$ state; e.g. for $n=4, i=2$:\\ 
\\
$|v^{(2)}_0\rangle=\frac{|1100\rangle+|1010\rangle+|1001\rangle}{\sqrt{3}}$\\
$|v^{(2)}_1\rangle=\frac{|1100\rangle+|0110\rangle+|0101\rangle}{\sqrt{3}},$\\
$|v^{(2)}_2\rangle=\frac{|1010\rangle+|0110\rangle+|0011\rangle}{\sqrt{3}},$\\
$|v^{(2)}_3\rangle=\frac{|1001\rangle+|0101\rangle+|0011\rangle}{\sqrt{3}}.$

\end{enumerate}

The $SWAP^{(n)}$ action on the $\big\{|v^{(i)}_k \rangle\big\}$ basis is isomorphic to the $SWAP^{(n)}$ action on the computational basis of $V_{i-1}$, where the isomorphism is the map taking $v^{(i)}_{s_k}$ to the computational state-vector with Hamming weight $i-1$ and qubits in positions given by the elements of $s_k$, in the $|1\rangle$ state. Therefore $V_{i,i}^\bot$ can be decomposed to irreducible invariant subspaces in the same way as $V_{i-1}$, by regarding the state-vectors $\big\{|v^{(i)}_k \rangle\big\}$ as the new basis for $V_{i,i}^\bot$.
\\
\\
\(V_{i,i}\) can be found by taking the orthogonal complement of $V_{i,i}^\bot$ in \(V_i\).

\end{enumerate}
This procedure is implemented as a recursive method on Mathematica. The code is available upon request from the authors.

\bibliographystyle{apsrev4-1}
\bibliography{references_2}

\end{document}